# Is "*L*-valine ferric chloride" a new nonlinear optical material?

Bikshandarkoil R. Srinivasan[1], Zbigniew Tylczyński[2]

[1]Department of Chemistry, Goa University, Goa 403206, India, Email: srini@unigoa.ac.in
[2]Faculty of Physics, Adam Mickiewicz University, Umultowska 85,
61-614 Poznań, Poland, Email: zbigtyl@amu.edu.pl

**Abstract**

We argue that "*L*-valine ferric chloride" crystal reported by Geetha *et al* [Curr. Appl. Phys. **15** (2015) 201-207] is not a new nonlinear optical material but instead a dubious crystal.

**Keywords:** *L*-valine ferric chloride; crystal growth; *L*-valine; nonlinear optical; dubious crystal.

**Comment**

The authors of a recent paper (title paper hereinafter) report to have grown a so called "*L*-valine ferric chloride" (**1**) crystal by the slow evaporation of an aqueous solution containing *L*-valine (*L*-val) and ferric chloride in 1:1 ratio [1]. Despite the claims of single crystal and powder X-ray diffraction, infrared and UV-visible spectra and thermal studies, the authors have not assigned any chemical formula for **1** excepting an unusual code LVFC for the so called new semi organic nonlinear optical crystal. In recent papers we have shown that use of abbreviations or codes as alternates for molecular formula of new compounds is an unscientific method of product characterization [2, 3]. In this comment we show that the authors have characterized the title crystal not based on valid scientific proof but based on an incorrect assumption that mixing of *L*-valine and ferric chloride in 1:1 ratio will result in the formation of **1**.

In the discussion of single crystal X-ray diffraction analysis the authors reported *"The axial length of unit cells are with a = 24.38 Å, b = 24.38 Å, c = 24.38 Å, the inter axial angles are α = β = γ = 90° and volume of the grown crystal is V = 14485 Å³. LVFC crystallise in Cubic P system"*. In addition to an unusual way of reporting a cubic system (in this system there are 15 space groups with primitive unit cell), it is noted that neither the temperature of cell measurement nor any ESD' values for the cell parameters are given. The volume of unit cell of **1** is incredibly large, almost 24 times more than the unit cell of *L*-valine which is one of the reactants used for the crystal growth. Although the cell appears to differ from that of *L*-valine [4], this cannot be considered as an acceptable scientific proof for the formation of a new compound. It is well documented that formulating new crystals based only on unit cell leads to erroneous conclusions [5]. We do not consider this experiment as acceptable single crystal X-ray diffraction result as the authors have not reported any refinement parameters and a CIF file to substantiate their work.

Under the heading 'powder X-ray diffraction the authors have mentioned, "*The sample was scanned in the range between 10 and 100°C. Fig. 2 represents the indexed powder diffractogram for the grown crystal of L-Valine Ferric Chloride. The sharp intensity peaks found in spectra shows good crystalline nature and purity of the grown crystal*". Since, the reported powder pattern (not spectrum) in the 2θ range 10 to 100° (not °C) shows a few spurious unindexed lines amidst lot of noise, we do not agree that such a pattern is a proof for the purity of the crystal. In our opinion such a poor quality powder pattern should not be used to make a scientific conclusion.

In an earlier paper we have shown that a so called *L*-valine zinc sulphate is actually *L*-valine based on the coincidence of both infrared spectra [6]. For the title crystal, authors reported infrared absorption signals to an accuracy of 0.01 cm$^{-1}$ and made assignments which are both questionable and incorrect. For example a band at 3377.36 cm$^{-1}$ assigned for OH$^{-}$ vibration indicates that *L*-valine does not exist as a zwitter ion in **1**. Although the authors report, "*The presence of other functional groups is verified from their respective absorption bands and they are in good agreement with those found in some complex amino acids*", we do not agree with this because it is inappropriate to make such conclusions based on a single spectrum. In this case, the IR spectrum matches with the spectrum of pure *L*-valine [6] indicating that the sample under study is the unreacted amino acid and not any so called "*L*-valine ferric chloride". Moreover, there are no bonds connecting Fe and Cl ions with C and H ones.

The authors write *"The crystal has very low absorption in the entire visible and NIR region. From 400 to 600 nm the absorption is absence its shows the material exhibiting NLO property"*. But from the reported fig.1 in the title paper, one can see that this crystal is dark yellow almost brown in colour. The dubious nature of this brown crystal can be evidenced from its optical spectrum which according to authors shows no absorption in the entire visible region.

The authors used LCR HIOKI HITESTER 3532-50 meter to investigate the dielectric properties and write:

- *"The material is characterized to load a resonant cavity and the sample permittivity is evaluated from the shift of the resonant frequency value compared to that of the empty cavity"*. This sentence is meaningless because this device measures the capacitance of the capacitor filled with the tested substance.
- *"The variations of the dielectric constant and dielectric loss as a function of frequency and dielectric constant are show in Fig. 9"*. In this figure we can see only the values of dielectric constant.

- *"The dielectric constant and dielectric loss increase with increase of temperature"*. In fig.9 of the title paper the dielectric constant is almost temperature independent in the range 50-90 °C.
- *"The characteristic of low dielectric constant with high frequency for a given sample suggests that the sample possess optical quality"*. The dielectric constant is very high, about 350, at the highest frequency 5 MHz.
- *"The metals are capable of decreasing the capacitance; the metal complex has a lower dielectric constant"*. This claim is unscientific.

In view of the above mentioned contradictory experimental data, namely an infrared spectrum matching with that of pure *L*-valine, a very large unexplained unit cell volume compared to that of *L*-valine, a highly questionable powder pattern, a brown coloured solid showing no absorption in the visible region and inappropriate dielectric data the so called "*L*-valine ferric chloride" cannot be considered as a new semi organic material but should be declared as a dubious crystal. A list of several dubious NLO crystal based on *L*-valine has been given recently [2] and most of these were characterized on similar lines. The so called "*L*-valine ferric chloride" is a new addition to the list of improperly characterized amino acid based crystal.

**In summary**, we have shown that a so called *L*-valine ferric chloride" is not a new nonlinear optical material but instead a dubious crystal.